\RequirePackage{fix-cm}
\documentclass[twocolumn,epjc3]{svjour3}  
\smartqed  % flush right qed marks, e.g. at end of proof
\RequirePackage{graphicx}
\usepackage{color}
\usepackage{amssymb,bbold}
\usepackage{graphicx,color}
\usepackage[usenames,dvipsnames]{xcolor}
\usepackage[section]{placeins}
\usepackage{amsmath}
\usepackage{subfigure}
\usepackage[normalem]{ulem}
\usepackage{booktabs}
\usepackage{xcolor}
\usepackage{epsfig,graphics,color,graphicx,amsmath}
\colorlet{darkgreen}{green!50!black}
\colorlet{brightyellow}{yellow!75!red}
\colorlet{orange}{red!50!yellow}
\colorlet{darkblue}{blue!60!black}
\colorlet{darkred}{red!80!black}
\def\bwt{\begin{widetext}}
\def\ewt{\end{widetext}}
\usepackage{soul}

\usepackage{mathtools}
\usepackage{mathrsfs}
\usepackage{bm}
\usepackage{relsize}
\usepackage{indentfirst}

\usepackage{xcolor}
\usepackage{slashed}

\usepackage{amssymb,amsmath,multirow,epsfig,graphicx,color}

\usepackage{epsfig}
\usepackage{graphicx,amsmath}
\def\be{\begin{eqnarray} &&}

\def\ee{\end{eqnarray}}

\begin{document}

\title{On the conformal limit of a QED-inspired model}

\author{O Oliveira \thanksref{e1,addr1} \and T Frederico \thanksref{e2,addr2} \and W. de Paula \thanksref{e3,addr2}}
\institute{CFisUC, Departament of Physics, University of Coimbra, 3004-516 Coimbra, Portugal \label{addr1} \and
Instituto Tecnol\'ogico de Aeron\'autica,  DCTA,
12228-900 S\~ao Jos\'e dos Campos,~Brazil \label{addr2}}
\thankstext{e1}{e-mail: orlando@uc.pt} 
\thankstext{e2}{e-mail: tobias@ita.br}
\thankstext{e3}{e-mail: wayne@ita.br} 
%\date{\today}

\maketitle

\begin{abstract}
A conformal invariant QED-inspired model is solved for a general covariant linear gauge using the Dyson-Schwinger equations for the propagators assuming a pure
vector like interaction.
The leading corrections to the asymptotic solutions and the exponents, that characterize the corrections to each of the two fermion propagator functions, are
computed as a function of the coupling and  gauge fixing parameter $\xi$. 
For the scalar component of the fermion propagator our findings generalizes for linear covariant gauges previous results found in the literature
and reproduce the outcome of the perturbative analysis of quenched QED in the Landau gauge.
Our solution for the exponent associated with vector component of the fermion propagator is new and, in the weak coupling regime, agrees with the estimation based 
on the perturbative analysis of quenched QED. 
Of the two critical exponents describing the conformal limit of the vector interaction, one of them is, in QED, associated with the regime where chiral symmetry 
is broken dynamically, which demands one mass scale, namely the Miransky scaling. 
A second mass scale has to be introduced at larger coupling constants and is associated with a change on the nature of the fermion wave function. 
This provides one example, that it is possible to find two interwoven cycles in Quantum Field Theory, albeit in a truncated framework, as it is known in the quantum few-body problem in the limit of a zero-range interaction.
\end{abstract}

\maketitle

%-------------------------------------------------------------------
%-------------------------------------------------------------------
\section{Introduction}\label{Sec:intr}

The combination of Quantum Mechanics and Special Relativity is Quantum Field Theory (QFT). In QFT the theories are summarized in a Lagrangian density $\mathcal{L}$. 
Moreover, from this function the equations of motion are derived, the generating functional for the Green functions is built that, hopefully, allow to solve the theory. 
The standard textbook solution for QFT is perturbation theory but there are other ways to access the dynamical content of a quantum field theory.
The infinite tower of Dyson-Schwinger equations (DSE) relates all the Green functions and provides a formal solution for a given QFT.
However, being an infinite tower of integral equations only a truncated version of the DSE, that necessarily ignores
the contribution of some of the Green functions and/or requires modeling others, can be solved.

The aim of the current work is to investigate the asymptotic behavior of the DSE for a certain type of theory that can be seen as a 
truncated version of QED. 
Instead of considering the complex dynamics of the full theory, the fermion-gauge vertex (one-particle irreducible Green function)
is approximate by its tree level value $\Gamma^\mu = \gamma^\mu$. Once defined the vertex function, the DSE for the
two point correlation functions form a closed set of integral equations whose solutions can be looked for. 
In a sense, this ``truncated theory'' can be viewed as an approximation of QED valid for the high energy regime. 
Of course there are issues related to gauge invariance, gauge covariance and multiplicative renormalization that should be considered when
solving the full theory,  see e. g.~\cite{Roberts:1994dr} and references therein for a review, that are not fully taken into account in the current analysis.
 
The family of theories where one considers a vector like interaction is widely used as an effective theory to describe many physical systems. 
In hadronic physics, for example, this is the case of the successful Maris-Tandy model \cite{Maris:1999nt}.
Moreover, for QCD, due to asymptotic freedom, in the high energy regime the quark-gluon vertex is dominated by a $\Gamma^\mu \approx \gamma^\mu$,  
module the color structure. This approximation is commonly used in hadronic physics and is the so called rainbow-ladder approximation.
Another example of a vector interaction as an effective description is the description of the electron dynamics at low energies 
in graphene. Indeed, for graphene or to certain classes of two-dimensional materials, a vector interaction can be associated either to the 
interaction of the electron with crystalline structure of the ionic background, to an interaction with the electromagnetic field or to model certain
types of defects, see e.g. the review \cite{CastroNeto:2007fxn}.

In the current manuscript we show that the QED-inspired model has two critical exponents in the ultraviolet limit. Our analysis generalizes the work of
Ref. \cite{Fukuda:1976zb} for the Landau gauge in two ways. It generalizes the results for any linear covariant gauge and identifies a second 
critical exponent that, within the approximation discussed, is also gauge dependent. The effective theory has two regimes of critical exponents. 
The configuration resembles the appearance of interwoven limit cycles, beyond the Efimov one~\cite{Efimov:1970zz}, obtained in quantum few-body problem in the limit of a zero-range interaction~\cite{Hadizadeh:2011qj,DePaula:2019ryz,Frederico:2019bnm}.
The one regime case has been extensively discussed in Ref.~\cite{Kaplan:2009kr} where it was explored its relation to the phenomenon of conformal breaking in different contexts.
 
In following we use the notation set in \cite{Oliveira:2022bar}, where the reader can find details on the derivations of the fundamental equations.

%==========================================================================
\section{The Dyson-Schwinger Equations \label{Sec:def}}

In Minkowski space-time, the fermion gap equation reads
\begin{eqnarray}
  S^{-1} (p)  &=& \left(  \slashed{p} - m \right) 
     -  \, i \, g^2 \, \int \frac{d^4 k}{( 2 \, \pi )^4}  ~  D_{\mu\nu} (k) ~\times \nonumber \\
     & & 
     \times
     \Big[ \gamma^\mu ~ S(p - k) ~   {\Gamma}^\nu (p-k, - p; k)  \Big] \ , 
     \label{DSE-FermionGap}
\end{eqnarray}
where
\begin{equation}
   S^{-1} (p) = A(p^2) \, \slashed{p} - B(p^2) + i \, \epsilon
   \label{FermionProp}
\end{equation}
is the inverse of the fermion propagator in momentum space,
\begin{eqnarray}
  D_{\mu\nu} (k)&=& - \, P^\perp_{\mu\nu}(k) \, D(k^2) - \frac{\xi}{k^2} \, P^L_{\mu\nu}(k)  \nonumber \\
  &=& - \left( g_{\mu\nu} - \frac{k_\mu k_\nu}{k^2} \right) \, D(k^2)  - \xi \, \frac{k_\mu k_\nu}{k^4}  \ , 
  \label{photonpropmom}
\end{eqnarray}
is the photon propagator in momentum space, in a general covariant gauge defined by the gauge fixing parameter $\xi$
and ${\Gamma}^\mu (p^\prime,  p; k)$ is the photon-fermion vertex where the convention is that all momenta are incoming and
$p^\prime +  p + k = 0$.

In our effective description of the interaction, the vertex will be approximate by its tree level perturbative expression. For
QED and for QCD, it is well known that this approximation does not comply with many requirements of the
theory. However, despite its limitations, we aim to build an effective model whose dynamics can be accessed from
the equations to the propagator and will use QED as an inspiration to define such an effective model.
For QED, the photon gap equation is
\begin{eqnarray}
  &&\frac{1}{D(k^2)}  = 
   k^2 
  - i \, \frac{g^2}{3} \,  \int \frac{d^4 p}{(2 \, \pi)^4} ~  \\
  & & \qquad\quad
   {Tr} \Big[ \gamma_\mu \, S(p) \, 
  {\Gamma}^\mu(p, -p + k; -k ) \, S(p - k ) \Big] \ . \nonumber 
     \label{DSE-PhotonGAP}
\end{eqnarray}
From the above DSE defined in Minkowski space-time, the corresponding Euclidean version can be obtained in the standard way.
Making the approximation $\Gamma^\mu = \gamma^\mu$, the Euclidean renormalized fermion equation become
\begin{multline}
B(p^2)  =  Z_0 \, m  ~ +  ~   Z_2 \, g^2 \int \frac{d^4k}{(2 \, \pi)^4} ~  F\Big( (p-k)^2 \Big) ~ \\
   \Bigg\{   
   3 \, D(k^2)  \, B\left((p-k)^2\right)   \, + \, \frac{\xi}{k^2} \, B\left((p-k)^2\right)  \Bigg\}  \mathcal{R} ( k^2 )  \\ \qquad\quad =  Z_0 \, m  ~ +  ~   Z_2 \, g^2  \, \Sigma_s (p^2)\ , \label{DSE:BEuclTree}
   \end{multline}
   for the scalar self-energy and
   \begin{multline}
 p^2 A(p^2)   =   Z_2 \, p^2 ~  - ~   Z_2 \, g^2 \int \frac{d^4k}{(2 \, \pi)^4} \, F\Big( (p-k)^2 \Big)  \\
   \Bigg\{ 
   D(k^2) \,   A\left((p-k)^2\right)  \left( \,  3 \, ({k} {p}) - {p}^2 -  2 \,  \frac{ ({k} {p})^2}{k^2}  \right)  \\ \quad \,\,
  + \, \frac{\xi}{k^2}  
A\left((p-k)^2\right)   \left( 2 \,  \frac{ ({k} {p})^2}{k^2}  - ({k} {p}) - {p}^2 \right)     \Bigg\}   \mathcal{R} ( k^2 ) \\ \qquad\quad =
    Z_2 \, p^2 ~  - ~   Z_2 \, g^2 \, p^2 \, \Sigma_v(p^2)  \, .    \label{DSE:AEuclTree}
\end{multline}
The equation for the photon propagator is
\begin{multline}
 \frac{1}{D(k^2)}  =  Z_3 \, k^2   -  \frac{8}{3} Z_2 g^2  \int \frac{d^4p}{(2 \, \pi)^4}  F(p^2)  F( (p-k)^2)  \mathcal{R} ( p^2 ) \\
 \Bigg\{   {A\left(p^2\right)} \, {A\left( (p-k)^2 \right)} \Big(    {p}^2 - {k} {p}  \Big) + 2 \, {B\left(p^2\right)} \, {B\left( (p-k)^2 \right)} \, \Bigg\}    \\
  = Z_3 \, k^2 ~  - ~  Z_2 \, g^2 \, k^2 \, \Pi(k^2) \, .
      \label{DSE:DEuclTree}
\end{multline}
In Eqs.~(\ref{DSE:BEuclTree}) to (\ref{DSE:DEuclTree}) the function
$\mathcal{R} ( k^2 ) $ is a regulator that makes the theory finite, $Z_0$, $Z_2$ and $Z_3$ are the renormalization constants, see
\cite{Oliveira:2022bar} for definitions, related to the renormalization of the mass, the fermion field, the gauge field and
\begin{equation}
   F(p^2) = \frac{1}{A^2(p^2) ~ p^2 + B^2(p^2) } \  .
   \label{Eq:FEucl}
\end{equation}
The dynamical model is defined in the Euclidean region by Eqs (\ref{DSE:BEuclTree}), (\ref{DSE:AEuclTree}), (\ref{DSE:DEuclTree}) and the regulator
$\mathcal{R}$.

In QED, the renormalization constants of the model can be determined from the conditions
\begin{equation}
    A(\mu_F^2) = 1 ,  \qquad  B(\mu_F^2) = m \quad\text{and}\quad D(\mu_B^2) = \mu_B^2 \, ,
\end{equation}
where $\mu_F$ and $\mu_B$ are the renormalization scales associated with the fermion and the boson field, respectively.
Then, the renormalization constants are
\begin{eqnarray}
    Z_0 &=& 1  - {Z_2 \over m } g^2 \, \Sigma_s (\mu_F^2) \, ,  \\
    Z_2 &=& {1 \over 1 - g^2 \, \Sigma_v (\mu_F^2)} \, , \\
    Z_3 &=& 1  + Z_2 \, g^2 \, \Pi(\mu_B^2) \, ,
\end{eqnarray}
and the renormalized equations DSE can be written as
\begin{eqnarray}
B( p^2) & = &   m  ~ - ~ { g^2  \over 1 - g^2 \, \Sigma_v (\mu_F^2) }  \,  \Sigma_s (\mu_F^2)    \nonumber \\
               & &   +  ~   { g^2 \over 1 - g^2 \, \Sigma_v (\mu_F^2)}  \, \Bigg( \Sigma_s (p^2)  - \Sigma_s (\mu^2_F)  \Bigg) , \label{RBp20} \\
 A(p^2) & = & {  1 - g^2 \, \Sigma_v (p^2)  \over  1 - g^2 \, \Sigma_v (\mu_F^2) }    \nonumber \\
       & = & 1 -  {g^2 \over  1 - g^2 \, \Sigma_v (\mu_F^2) }  \! \Bigg( \Sigma_v (p^2) - \Sigma_v (\mu^2_F)  \Bigg)  , \label{RAp20}  \\
 \frac{1}{D(k^2)} & = & k^2 ~\times \nonumber \\
 & & \!\!\!\!\!\!\!\!
   \times \Bigg( 1 - { g^2 \over 1 - g^2 \, \Sigma_v(\mu_F^2) } \Big( \Pi(k^2) - \Pi(\mu_B^2) \Big) \Bigg)  . \label{DSE:D1250}
\end{eqnarray}
If one uses a hard cutoff $\Lambda$ as a regulator for the computation of the self energies 
$\Sigma_{s, v}$ and of $\Pi$, the vector component of the fermion self energies $\Sigma_v (\mu_F^2)$ and $\Pi (\mu^2_B)$ diverge
as  $\Lambda \rightarrow \infty$.
In principle, at least in QED, this divergence is taken care by the dynamics of the theory. 
Indeed, in perturbation theory and to lowest order in the coupling constant the equations for the fermions are finite. They are given by
\begin{eqnarray}
B( p^2) & = &   m  ~ +  ~   g^2  \, \Bigg( \Sigma_s (p^2) - \Sigma_s (\mu^2_F) \Bigg) \ , \label{RBp2Pert} \\
 A(p^2) & = & 1 ~ - ~ g^2 \Bigg( \Sigma_v (p^2) - \Sigma_v (\mu^2_F)  \Bigg)  , \label{RAp2Pert}  \\
 \frac{1}{D(k^2)} & = & k^2 ~ \Bigg( 1 ~ - ~   g^2 \,  \Big( \Pi(k^2) - \Pi(\mu_B^2) \Big) \Bigg)  ,\label{DSE:D125Pert}
\end{eqnarray}
and the subtractions cancel the divergent parts for the first two equations. In the simple approximation used for the vertex, this subtraction
is not enough to render the photon equation finite. As will be seen later, the photon propagator equation requires further subtractions.
Let's ignore for the momentum this issue and proceed to arrive at an effective approximation to QED.
From the point of view of building an effective and truncated version of QED, in the spirit of \cite{Weinberg:1996kr},
the divergences associated with $\Sigma_v$
can be absorbed in the definition of the physical mass $m_{ph}$ and the physical coupling constant $g_{ph}$ that are taken as
\begin{equation}
   m_{ph} = m  ~ - ~ { g^2  \over 1 - g^2 \, \Sigma_v (\mu_F^2) }  \,  \Sigma_s (\mu_F^2) 
\end{equation}
and
\begin{equation}
   g^2_{ph} =  { g^2  \over 1 - g^2 \, \Sigma_v (\mu_F^2) } \ .
   \label{effective_coupling}
\end{equation}
Then, the renormalized DSE can be written as
\begin{eqnarray}
B( p^2) & = &   m_{ph}   ~  +  ~    g^2_{ph}   \Bigg(  \Sigma_s (p^2) - \Sigma_s (\mu^2_F)  \Bigg) , \label{RBp2} \\
 A(p^2) & = & 1 ~ - ~  g^2 _{ph} \Bigg( \Sigma_v (p^2) - \Sigma_v (\mu^2_F)  \Bigg)  , \label{RAp2}  \\
 \frac{1}{D(k^2)} & = & k^2 ~ \Bigg( 1 ~ - ~  g^2_{ph}  \Big( \Pi(k^2) - \Pi(\mu_B^2) \Big) \Bigg)  . \label{DSE:D125}
\end{eqnarray}
These equations define our effective model and will be analysed in the asymptotic regime.
Furthermore, as discussed below, due to the divergences
in the photon equation, the photon propagator will be freezed to its tree level value and, therefore, our analyses refers to the quenched
theory.

%================================================================================

\section{Conformal Invariance and the UV limit}

The dynamical model defined by the DSE (\ref{RBp2}) -- (\ref{DSE:D125}) has a single reference scale,
the fermion physical mass. The solutions of the model are characterized by $m$ and by the regulator used to render the theory
finite.  
For an energy regime where the fermion mass is small enough, i.e. in the UV limit, or in the chiral limit, the above set of integral equations 
becomes conformal invariant. In the conformal limit the solutions for $B(p^2)$, $A(p^2)$ and $D(p^2)$ are parametrized by exponents 
describing their asymptotic behavior and to investigate this regime the mass term can be ignored. Let us consider the ansatze
\begin{eqnarray}
B(k^2) & = &   \lim_{\epsilon \to 0^-}  \left( b_1 \, \left( k^2 \right)^\epsilon+ b_2 \, \left( k^2 \right)^\eta\right)  \, , \label{ABAnsatzB}
 \\
A(k^2) & = & \lim_{\epsilon \to 0^-}  \left(a_1 \, \left( k^2 \right)^\epsilon+ a_2 \, \left( k^2 \right)^\gamma \right)\, , \label{ABAnsatzA}
\\
D(k^2)^{-1} &=& k^2 \, \left( d_1 + d_2 \, \left( k^2 \right)^ \delta \right )
\label{ABAnsatzD}
\end{eqnarray}
with Eqs.(\ref{RBp2}) and (\ref{RAp2}) written as:
\begin{eqnarray}
B( k^2)  &=&    \lim_{\epsilon \to 0^-}  \Big[ m_{ph}(\zeta +(1-\zeta)b_3^{\epsilon})(k^2)^\epsilon \nonumber \\ && \hspace{1cm}~ + ~   g^2  \, \Big( \Sigma_s (k^2) - \Sigma_s (\mu^2_F) \Big) \Big] \ , \label{RBp2Pert2} \\ 
 A(k^2)  &=&  \lim_{\epsilon \to 0^-}\Big[ (\overline\zeta+(1-\overline\zeta)a_3^\epsilon)(k^2)^\epsilon \nonumber \\&&\hspace{1cm}~ - ~ g^2 \Big( \Sigma_v (k^2) - \Sigma_v (\mu^2_F)  \Big) \Big] \,, \label{RAp2Pert2} 
\end{eqnarray}
in order to coop with the ansatze made in Eqs.~\eqref{ABAnsatzB} and~\eqref{ABAnsatzA}.
The parameters $\zeta$, $\overline\zeta$, $a_i$, $b_i$ and $d_i$ are constants and the limit is calculated at the end of the analysis. The above modification of
the equations of motion was introduce to handle the 
$\log$ terms that naturally appear in
a perturbative approach and that break conformal invariance. Indeed, in this way, by a proper choice of $\zeta$, $\overline\zeta$, $a_i$ and $b_i$ conformal invariance is recovered.

The exponents $\eta$, $\gamma$ and $\delta$ are, in principle, negative numbers that parametrize the
solutions of the model in the conformal limit.
The constants $b_2$, $a_2$, $d_2$ are  not dimensionless but have dimensions given by positive powers of mass.
The ansatze for the various functions assumes that the tree level perturbative solution should be recovered in the high energy regime.

Assuming that the propagators functions are as in (\ref{ABAnsatzB}) to (\ref{ABAnsatzD}), it is possible to compute the self energies; 
see App. \ref{App:Self} for details. The fermion self energies are
\begin{eqnarray}
\Sigma_s (p^2) & = &- 
~{b_1 \over a_1^{2} }\,   {( 3 \, d_1^{-1}  \, + \, \xi ) \over 16 \, \pi^2 \, \epsilon (1+\epsilon)} \,  \left( p^2 \right)^ \epsilon \ ,
 \nonumber \\
  & &  \qquad 
  - ~{b_2 \over a_1^{2} }\,   {( 3 \, d_1^{-1}  \, + \, \xi ) \over 16 \, \pi^2 \, \eta(1+\eta)} \,  \left( p^2 \right)^ \eta \ ,
  \label{Eq:Sigma_s}
\end{eqnarray} 
\begin{eqnarray}
\Sigma_{v}(p^2)  & = &   ~ { 1 \over a_1 }  \, {\xi\over 16 \pi^2 \, \epsilon (1+\epsilon)}  \,  \left( p^2 \right)^\epsilon  \, \nonumber \\
   & &   \qquad
        +  ~ { a_2 \over a_1^{2} }  \, {\xi\over 8 \pi^2 \, \gamma (2+\gamma)}  \,  \left( p^2 \right)^\gamma \ .
          \label{Eq:Sigma_v}
\end{eqnarray}
with $-1< \eta < 0$ and $-1<\gamma<0$ for the integrations to be UV and IR finite. 

%Then, the equations for the fermion become

Taking into account Eqs.~\eqref{Eq:Sigma_s} and ~\eqref{Eq:Sigma_v}, and considering that $\mu_F\to\infty$ and $\epsilon,\gamma, \eta<0$
then $\Sigma_{s,v}(\mu^2_F)\to0$:  
\begin{multline}
b_1 \, \left( p^2 \right)^\epsilon + b_2 \, \left( p^2 \right)^\eta  =  
\\
=m_{ph} (\zeta +(1-\zeta)b_3^{\epsilon}) (p^2)^\epsilon + g^2_{ph}  \Sigma_s(p^2)  \, ,\\ \\
a_1 \, \left( p^2 \right)^\epsilon + a_2 \, \left( p^2 \right)^\gamma  = \hspace{3cm} \\ =(\overline\zeta+(1-\overline\zeta)a_3^\epsilon)(p^2)^\epsilon- g^2_{ph}  \Sigma_v(p^2)  \, .\label{subtraction20}
\end{multline}
Then introducing Eqs.~\eqref{DSE:AEuclTree21} and \eqref{DSE:AEuclTree22} in the equations above one finds that:
\begin{multline}
b_1  \left( p^2 \right)^\epsilon + b_2  \left( p^2 \right)^\eta  =   m_{ph} (\zeta +(1-\zeta)b_3^{\epsilon})(p^2)^\epsilon \\ - g^2_{ph} b_1 p^{2 \epsilon} \, a_1^{-2} \,   {( 3 \, d_1^{-1}  \, + \, \xi ) \over 16 \, \pi^2 \, \epsilon(1+\epsilon)} \\ 
 -g^2_{ph} b_2 p^{2 \eta} \, a_1^{-2} \,   {( 3 \, d_1^{-1}  \, + \, \xi ) \over 16 \, \pi^2 \, \eta(1+\eta)}  \,, \end{multline}
 and
 \begin{multline}
a_1 \, \left( p^2 \right)^\epsilon + a_2 \, \left( p^2 \right)^\gamma  =  (\overline\zeta+(1-\overline\zeta)a_3^\epsilon)(p^2)^\epsilon \\- g^2_{ph} a_1^{-1}   \, \xi \, p^{2\epsilon } {1\over 8 \pi^2 \, \epsilon (2+\epsilon)}\\
- g^2_{ph}\,a_1^{-2} \, a_2  \, \xi \, p^{2\gamma } {1\over 8 \pi^2 \, \gamma (2+\gamma)} \,.
\label{subtraction21}
\end{multline}
We can choose the constants $a_3$ and $b_3$ in order to cancel the $1/\epsilon$ terms. Then, equating the $p^{2\epsilon}$ terms we find that
\begin{equation}
    b_1=m_{ph}\zeta\quad\text{and}\quad a_1=\overline\zeta\, .
\end{equation}
Next equating the $p^{2\gamma}$ terms, we get that:
\begin{equation}
    1= -g^2_{ph}   \, a_1^{-2} \,   {( 3 \, d_1^{-1}  \, + \, \xi ) \over 16 \, \pi^2 \, \eta(1+\eta)} \,,
\label{Eqa2}
\end{equation}
and
\begin{equation}
    1  = 
- g^2_{ph}\,a_1^{-2}  \, \xi \,  {1\over 8 \pi^2 \, \gamma (2+\gamma)} \,.
\label{Eqb2}
\end{equation}

The coupling constant can be read from
 (\ref{Eqa2}) and (\ref{Eqb2}) that imply 
 \begin{equation}
 \alpha  =  {g^2_{ph} \over 4 \, \pi a_1^2 }   =  { - 4 \, \pi \, \eta \, ( 1 + \eta ) \over 3 \, d_1^{-1}  + \xi } 
  = 
 { - \, 2 \, \pi \, \gamma \, ( 2 + \gamma )  \over \xi } \,, \label{alphas} 
\end{equation}
and relates the exponents $\eta$ and $\gamma$ with $\xi$ and with the asymptotic value of the photon propagator. Note that for $\xi = 0$ the last equality is not applicable.
This relation is independent of the value of the coupling constant $\alpha$
and requires 
\begin{equation}
   \gamma \, ( 2 + \gamma ) = 2 \, {\xi \over 3 \, d_1^{-1} + \xi } \, \eta \, ( 1 + \eta ) 
   \quad\text{ for }\quad \xi \ne 0 \ . \label{BApowers}
\end{equation}

%============================================================
\subsection{The gauge boson DSE in the asymptotic regime}

From the point of view of QED-inspired model, the analysis of the vector boson equation for the propagator, i.e. Eq. (\ref{DSE:D125Pert}),  
assuming  that (\ref{ABAnsatzB}) to (\ref{ABAnsatzD}) hold for all momenta encounters UV and IR divergences. 
The divergences can be identified via naive power counting analysis and are associated only with the contribution coming from
the $A(p^2) A((p-k)^2)$ term. In QED theory these divergences are cured by taking into account its complete dynamics.
However, in the (truncated) effective QED-inspired model these divergences have to be cured
to have a finite theory.

One way to make the (truncated) effective model finite considers, for example, a finite cutoff or dimensional regularization, or any other kind of regularization, in the integration over momentum.
Another solution is to separate 
\begin{equation}
\Pi (k^2) = \Pi (k^2; \Lambda) + \Pi_{\text{finite}} (k^2) \ ,
\label{photon:finite}
\end{equation}
where the first term has all the divergent terms and the second term is finite, and use in Eq.~(\ref{DSE:D125Pert}) only $\Pi_{\text{finite}}$ instead
of $\Pi$. The separation of $\Pi$ into a divergent part and a finite term is not unambiguously and, in principle,
a physical motivated extra condition should be used to define $\Pi_{\text{finite}}$. Any condition that allows such a separation defines an
effective field theory model. Another solution to handle the problem of the divergences is to freeze the vector propagator to a given \textit{a priori}
function form. A ``natural'' choice is to set $D(k^2)$ to its tree level form, maybe including a small gauge boson mass to regulate possible IR
divergences. The ``quenched'' QED, where the fermion loop contribution to the photon dynamics is ignored, belongs to this later class of solutions.
From the point of view of building an effective theory, any of the above solutions is allowed.

Let us assume that the separation of the gauge boson self energy as in (\ref{photon:finite}) is implemented. In this case one can relate 
the gauge boson exponent $\delta$ to $\eta$ and/or $\gamma$. The analysis of the corresponding regularised  gauge boson DSE shows
that there are regimes where $\delta$ is connected to either of the fermionic exponents. From the point of view of the DSE it is not clear
which of the regimes will take place.

%==================================================================
\subsection{Power law exponents}

The exponents $\eta$ and $\gamma$ that characterize the conformal solution of the effective model
can be computed assuming that the photon propagator takes its tree level value, i.e that $d_1 = 1$. Then, Eq. (\ref{alphas}) that gives $\eta$ and $\gamma$ in terms of $\alpha$ and of $\xi$ becomes
\begin{equation}
 \alpha   =   { - \, 4 \, \pi \, \eta \, ( 1 + \eta )  \over 3 + \xi }  =   { - \, 2 \, \pi \, \gamma \, ( 2 + \gamma ) \, \over \xi }  \ .
\end{equation}
The mathematical solutions of these equations are
\begin{equation}
\eta = - \, \frac{1}{2} \pm \frac{1}{2} \sqrt{ 1 - \frac{\alpha}{\pi} ( 3 + \xi )}
\label{etaX}
\end{equation}
and
\begin{equation}
\gamma = - \, 1 +  \sqrt{ 1 - \frac{\alpha}{2 \pi} \, \xi } \ .
\label{gammaX}
\end{equation}
However, the range of possible values for $\eta$ and $\gamma$ that makes the effective theory finite implies
that the conformal solution exists if and only if
\begin{equation}
    \alpha ~ <  ~ \text{Min}\left( \frac{2 \, \pi}{\xi}, \, \frac{\pi}{3 + \xi} \right)= \frac{\pi}{3 + \xi} \ .
    \label{alpha_critical}
\end{equation}    

\begin{figure}[t]
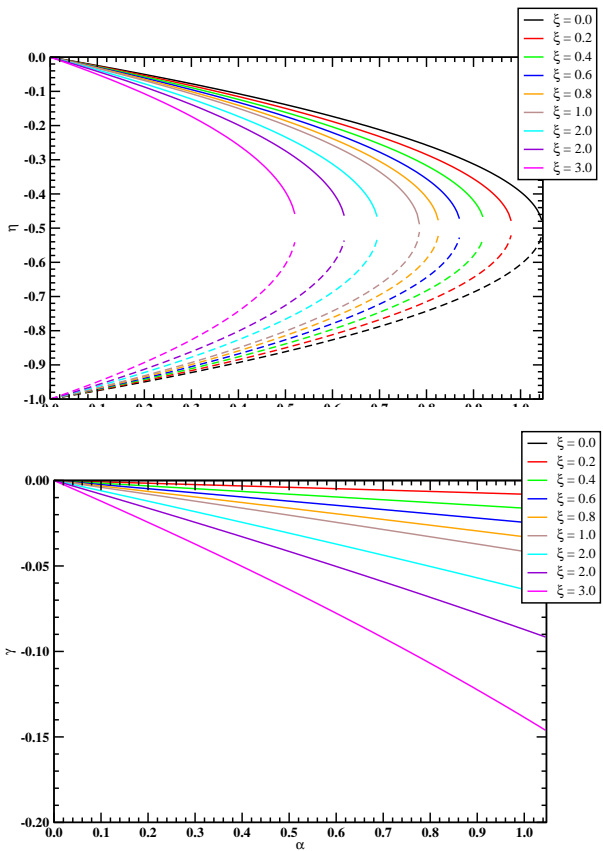
 
   \centering
   \includegraphics[scale=0.3]{expoentes.eps} 
   \\
   \includegraphics[scale=0.3]{expoentes_gamma.eps}
   \caption{The exponents $\eta$ and $\gamma$ as a function of the coupling as given in Eqs (\ref{etaX}) and (\ref{gammaX}), respectively.
   In the upper plot, the full line is solution in (\ref{etaX}) when one takes the plus sign, while the dashed lines are the solutions with the
   minus sign.}
   \label{fig:expoentes}
\end{figure}

The solutions of Eqs (\ref{etaX}) and (\ref{gammaX}) that are compatible with the bounds on $\eta$ and $\gamma$ discussed previously 
are plotted in Fig. \ref{fig:expoentes}. 
For the exponent $\eta$,  equation (\ref{etaX}) returns two solutions but that with the highest modulus of the exponent is subleading and,
therefore, the leading correction is given by the solution with plus sign in (\ref{etaX}). 
The results of Fig. \ref{fig:expoentes} suggest that the leading corrections to the asymptotic solution are milder for $A(p^2)$ than for $B(p^2)$,
in the sense that $\gamma$ is closer to zero than $\eta$. Their relative value depends on the value of the constants
$a_2$ and $b_2$ that we are not able to compute. In what concerns the dependence on the coupling constant $\alpha$, the
corrections to the asymptotic solution seem to decrease as $\alpha$ increases. There is also some dependence on the
$\xi$, with  the critical value for $\alpha$, see Eq. (\ref{alpha_critical}), decreasing when $\xi$ increases. 

Taking the effective model as an approximation to QED, the dependence of $\alpha_c$ with $\xi$ shows that a
$\alpha_c = 1/137$ occurs for $\xi = 137 \, \pi - 3 \approx 427.398\dots$. 
On the other hand to have a $\alpha_c = 0.3$, a typical value considered within a 
QCD inspired description, one should have $\xi = \pi/0.3 - 3 \approx 7.47\dots$
Another example is  the Yennie-Fried gauge in QED \cite{Fried:1958zz,Yennie:1961ad} where $\xi = 3$. For this gauge the
corresponding $\alpha_c = \pi / 6 \approx 0.52\dots$ We recall the reader that for QED in the Yennie-Fried gauge the theory is IR finite,
see e.g. \cite{Tomozawa:1979xn,Eides:2001dw} and references therein, at least in its lowest order approximation. For a review of the phenomenon of conformality breaking giving rise to different phases see Ref.~\cite{Kaplan:2009kr}.

The Euclidean space DSE given in (\ref{RBp2}) and (\ref{RAp2}) have been studied in quenched QED long time ago.
Indeed, the exponent derived in Eq.~(\ref{etaX}) is a generalization for linear covariant gauges of the result obtained in Ref.~\cite{Fukuda:1976zb}, where the Landau gauge was considered. The UV analysis in a fixed gauge was also performed in Ref.~\cite{Johnson:1973pd}. In the  literature for QED, this exponent is related with the possibility of having dynamical symmetry breaking for $\alpha > \alpha_c = \pi/(\xi + 3)$.
Furthermore, the strong dependence on the gauge parameter was investigated and
found that it is connected with the choice of such a simple vertex as $\gamma^\mu$.
Indeed, the recovering of gauge covariance and multiplicative renormalization, that demand a more complex vertex,
seems to wash out the dependence on $\xi$ as can be seen in e.g. \cite{Curtis:1990zr,Albino:2018ncl} and references therein.

In what concerns the exponent $\gamma$, in the literature, to the best knowledge of the authors, the closest solution to ours can be found in
\cite{Curtis:1991fb}, where the authors used perturbation theory with the DSE for a massless fermion
and a vertex with other tensor structures than the tree level vertex and arrived at $\gamma = - \, \alpha \, \xi / 4 \, \pi$. This 
results agrees with the solution  (\ref{gammaX}) in the weak coupling regime and in lowest order in the coupling constant $\alpha$. 
A major difference to the results of \cite{Curtis:1991fb} being that our description of the propagators predicts a second critical exponent. If the critical coupling constant associated with $\eta$ ($\alpha_c = \pi/(3 + \xi)$) 
occurs at smaller values of $\alpha$ and is associated with chiral symmetry breaking 
\cite{Miransky:1984ef,Miransky:1985wzx,Gusynin:1986fu,Miransky:1989qc,Bardeen:1990im,Braun:2010qs,Antipin:2012kc},
the critical exponent associated with $\gamma$  ($\alpha_c = 2\,  \pi/\xi$) does not appear in previous studies.
The critical couplings have, in both cases, is strongly dependent on the gauge parameter $\xi$ and, in principle, this dependence on
$\xi$ is reduced when more realistic vertices are considered. The main difference between the calculation described here and that of~\cite{Curtis:1991fb} being that the later use tree level propagators, while here we go beyond the tree level propagators and
include corrections that are described by power laws. In both cases the boson propagator is treated as massless.
The second critical exponent appears due to the corrections associated with the fermion propagators.

%=====================================================================
\section{The Beta Function}

The truncated model under discussion has an effective coupling that is defined by~(\ref{effective_coupling})
and, therefore, its dependence on the renormalization scale $\mu_F$ is known if $A$, $B$ and $D$ are
given by their asymptotic expressions; see Eqs.~(\ref{ABAnsatzB}) to (\ref{ABAnsatzD}). This allows to compute
exactly the $\beta$ function of the theory. For the Landau gauge where $\xi = 0$, $a_1 = 1$ and $a_2 = 0$ it turns
out that $\Sigma_v(\mu) = 0$ and, therefore,
\begin{equation}
   g^2_{ph} = g^2 \ ,
\end{equation}   
i.e. the coupling constant does not run unless the bare coupling constant runs. To proceed with the analysis of the
$\beta$ function let us write
\begin{equation}
   g^{-2}_{ph} = g^{-2} ( \mu ) + \Delta g^{-2} ( \Lambda, \xi, \mu ) \ ,
   \label{Eq:gg}
\end{equation}   
where $\Delta g^{-2}$ is defined to make $g^2_{ph}$ finite\footnote{Note that the definition of the physical effective coupling constant~(\ref{effective_coupling})
can be rewritten as $g^2_{ph} = 1 /( g^{-2}  - \Sigma_v (\mu^2_F))$. Note that with a convenient definition for $\Delta g^{-2}$ the result of Eq.~(\ref{Eq:gg}) applies to any linear covariant gauge.}. Then, for the Landau gauge
\begin{equation}
   g^{2}_{ph} = g^{2} ( \mu ) 
\end{equation}   
and, therefore, one can insert in the model any convenient running via this last relation\footnote{Although we are considering only the Landau gauge,
the procedure can be extended to any other $\xi$ value.}. Extrapolating our analysis to a QCD effective model, a possibility is incorporate the QCD running of the coupling constant that in perturbation
theory at one-loop approximation reads
\begin{equation}
  \alpha (\mu^2)  = \frac{ 4 \, \pi}{ \beta_0 \, \ln ( \mu^2 / \Lambda^2 ) }
\end{equation}
where $\beta_0 = 11 - 2 \, N_f / 3$, $N_f$ being the number of massless fermion degrees of freedom and $\Lambda$ is the QCD scale at which the
color interactions become strong.

We should comment that  within QCD, where  the quark-gluon vertices are enhanced in the infrared region~\cite{Oliveira:2020yac,Oliveira:2018ukh,Oliveira:2018fkj} and in the UV region asymptotic freedom is valid, the Miranski scaling will never show up in the solution on the  Dyson-Schwinger  equation for the dressed quark, and the UV behaviour of the pion Bethe-Salpeter amplitude will be dominated by the power law solution, and eventually impact the end-point behaviour of the associated parton distributions~\cite{Noronha:2023hjg,dePaula:2023ver,dePaula:2022pcb}.

%=====================================================================
\section{Summary and Conclusions}

We found that the conformal solution of the Dyson-Schwinger equations for the dressed fermion in the effective model with bare vertex and bare gauge boson has two critical exponents. Besides the power-law behaviour of the fermion scalar self-energy in the UV region, the fermion wave function also acquires a power law dependence. This solution presents a new second phase transition in the effective model associated with the power law behaviour of the fermion wave function turning to a log-periodic one, beyond the Miransky scaling solution when chiral symmetry is spontaneously broken. 

The second phase transition has a critical coupling depending on the gauge, which also appears in the first transition. In particular, at the Landau gauge only the standard Miransky scaling prevails. Assuming that the coupling is large enough, the Bethe-Salpeter amplitude associated with the massless Goldstsone boson, according to the axial vector Ward identities~\cite{Maris:1997hd}, will have, besides the log-periodicity in the pseudoscalar vertex, another one, carrying a new scale that has to be introduced in the model through the fermion wave function renormalization, which appears in the fermion propagator. Our findings provides another example,  within a truncated model of QFT, of interwoven cycles, beyond the Efimov one, which were already discussed in the context of the quantum mechanical few-body problem in the limit of a zero-range interaction.

{\it This work was partly supported by the FCT - Funda\c c\~ao para a Ci\^encia e a Tecnologia, I.P., under Projects Nos. UIDB/04564/2020, UIDP/04564/2020.
OO acknowledges financial support from  grant 2022/05328-3, from S\~ao Paulo Research Foundation (FAPESP). W. d. P. acknowledges the partial support of the National Council for Scientific and Technological Development (CNPq) under Grant No. 313030/2021-9 and the partial support of the Coordination for the Improvement of Higher Education Personnel (CAPES) under Grant No. 88881.309870/2018-01. T. F. thanks
the financial support of CNPq (Grant No.  306834/2022-7), CAPES (Finance Code 001) and FAPESP (Grant 
No.  2019/07767-1). This work is a part of the
project Instituto Nacional de  Ci\^{e}ncia e Tecnologia - F\'{\i}sica
Nuclear e Aplica\c{c}\~{o}es  Proc. No. 464898/2014-5.}

%===================================================================
\appendix

%===================================================================
\section{Fermion self energies}\label{App:Self}

Here we summarize the computation of the fermion self energies that enter in the DSE.

%===================================================================
\subsection{The scalar part of the fermion self energy}

The scalar part of the fermion self energy $ \Sigma_s(p^2)$ appears in the integral
equation for $B(p^2)$. Let us write $ \Sigma_s(p^2)  = \Sigma^{(1)}_{s}(p^2) + \Sigma^{(2)}_{s}(p^2) $
where
\begin{multline}
\Sigma^{(1)}_{s}(p^2)  = \\ =  b_1 \int \frac{d^4q\,q^{2\epsilon}}{(2 \, \pi)^4} \, F(q^2) \Bigg\{   
   D((q-p)^2) \,  3  \, + \, \frac{\xi}{(q-p)^2}   \Bigg\}  \, \\ =  b_1 \, a_1^{-2} \,  ( 3 \, d_1^{-1}  \, + \, \xi ) \int \frac{d^4q}{(2 \, \pi)^4} \, {1\over \, q^2}  \, {\, q^{2\epsilon} \over (q-p)^2} \\ =  - b_1 p^{2 \epsilon} \, a_1^{-2} \,   {( 3 \, d_1^{-1}  \, + \, \xi ) \over 16 \, \pi^2 \, \epsilon(1+\epsilon)} \, 
\end{multline} 
and
\begin{multline}
\Sigma^{(2)}_{s}(p^2) =  \\ = b_2   \int \frac{d^4q\,q^{2\eta}}{(2 \, \pi)^4}  F(q^2) \Bigg\{   
   D((q-p)^2)  3   +  \frac{\xi}{(q-p)^2}   \Bigg\}   \\ =  b_2 \, a_1^{-2} \,  ( 3 \, d_1^{-1}  \, + \, \xi ) \int \frac{d^4q}{(2 \, \pi)^4} \, {1\over \, q^2}  \, {q^{2\eta} \over (q-p)^2} \\ =  - b_2 p^{2 \eta} \, a_1^{-2} \,   {( 3 \, d_1^{-1}  \, + \, \xi ) \over 16 \, \pi^2 \, \eta(1+\eta)}
\end{multline}
and $-1< \eta < 0$ for convergency of the last integral. 

%=======================================================================
\subsection{The vector part of the fermion self energy}\label{AppVector}

Similarly, the vector part of the fermion self energy appears in the integral equation
associated with $A(p^2)$ and we write as 
$ \Sigma_v(p^2)  = \Sigma^{(1)}_{v}(p^2) + \Sigma^{(2)}_{v}(p^2) $
with
\begin{multline}
\Sigma^{(1)}_{v}(p^2)   =   p^{-2} a_1^{-1}  \int \frac{d^4q}{(2 \, \pi)^4} 
   {q^{2\epsilon} \over q^2 (q-p)^2}\\ \times \Bigg\{ 
    {1\over d_1} \, \left( 2 p^2 - 3 \, pq - \frac{2 \, (p^2 - pq)^2}{(q-p)^2}  \right)  
 \\ + \, \xi  
\left(  - 2 p^2 + pq + \frac{2 \, (p^2 - pq)^2}{(q-p)^2}  \right)  ~   \Bigg\} \,   ,\\
=  a_1^{-1}   \, \xi \, p^{2\epsilon } {1\over 8 \pi^2 \, \epsilon (2+\epsilon)}   \,  .
      \label{DSE:AEuclTree21}
\end{multline}
and
\begin{multline}
\Sigma^{(2)}_{v}(p^2)  =   p^{-2}  a_1^{-2}\,  a_2 \,  \int \frac{d^4q}{(2 \, \pi)^4}  
   {q^{2\gamma} \over q^2 (q-p)^2} \\ \times \Bigg\{ 
    {1\over d_1} \, \left( 2 p^2 - 3 \, pq - \frac{2 \, (p^2 - pq)^2}{(q-p)^2}  \right)  
 \\ + \xi  
\left(  - 2 p^2 + pq + \frac{2 \, (p^2 - pq)^2}{(q-p)^2}  \right)  ~   \Bigg\} \,  \\ =    a_1^{-2} \, a_2  \, \xi \, p^{2\gamma } {1\over 8 \pi^2 \, \gamma (2+\gamma)}  \, .
      \label{DSE:AEuclTree22}
\end{multline}
The second terms is defined only if $-1<\gamma<0$.

%=================================================

\end{document}